\documentclass[review]{ametsoc}

\journal{jpo}

\bibpunct{(}{)}{;}{a}{}{,}

\title{Multiple Equilibrium States of the Loop Current in the Gulf of Mexico}

\authors{Vitalii A. Sheremet}

\affiliation{Graduate School of Oceanography, University of Rhode Island, Narragansett, Rhode Island, USA}

\extraauthors{Arham Amin Khan}

\extraaffil{Department of Mechanical Engineering, University of Delaware, Newark, Delaware, USA}

\extraauthors{Joseph J. Kuehl\correspondingauthor{Joseph Kuehl, Department of Mechanical Engineering, University of Delaware, 130 Academy Street, Newark, DE 19716, USA.}}

\extraaffil{Department of Mechanical Engineering, University of Delaware, Newark, Delaware, USA}

\email{jkuehl@udel.edu}

\abstract{
It is known that western boundary currents, which encounter a gap in their supporting boundary, assume two dominant steady states: a loop current state and a gap leaping state, and that transitions between these states display hysteresis.  However, a question of whether the idealized geometries considered to date apply to the Gulf of Mexico Loop Current (LC) remained.  Here, the nonlinear potential vorticity advection-diffusions equations are solved, for Gulf of Mexico topography,  using Newton's Method.  We demonstrate that, in application to the LC in the Gulf of Mexico, the original conclusions do hold and additionally describe peculiarities of the more realistic steady states that have implications for the LC modeling and forecasting.
}

\begin{document}

\maketitle

%








\nolinenumbers

\section{Introduction}

Responding to the external forcing and intrinsic dynamics, the Loop Current (LC) in the Gulf of Mexico exhibits 
dramatic variations. In a retracted state, the LC leaps directly from the Straits of Yucatan near Cuba to the Florida Straits, and in an extended state, the LC penetrates far into the Gulf of Mexico and forms a large eddy which then may pinch off and drift westward \citep{donohue_gulf_2016, donohue_loop_2016, LPKKZ_JRM_2008} (Figure \ref{F1} upper right and left panels, respectively).  This process exerts a major control over the strength of hurricanes~\citep{kafatos_role_2006}, dispersion of pollutants~\citep{weisberg_movement_2017}, offshore energy operations~\citep{koch_gulf_1991}, and even coastal ecosystem health~\citep{Hetland_GRL_1999, weisberg_why_2014}.  Despite decades of scientific inquiry and major field program initiatives, the fundamental problem of LC predictability remains \citep{committee_on_advancing_understanding_of_gulf_of_mexico_loop_current_dynamics_understanding_2018}. \cite{sheremet_hysteresis_2001} originally discovered that an idealized western boundary current
with Munk profile running along a straight wall with a gap, for the same governing parameters, may be in two steady states: penetrating into the western basin due to the $\beta$-effect and leaping across the gap due to flow inertia.  Furthermore, the time variability of the current will involve a hysteresis, a dependence on prior evolution.  
These results were later confirmed in rotating table laboratory experiments for more general barotropic boundary currents~\citep{sheremet_gap-leaping_2007, kuehl_identification_2009} as well as for baroclinic boundary currents~\citep{kuehl_two-layer_2014, mcmahon_viscous_2020}. In addition, more advanced modeling studies have confirmed the existence of multiple steady states with hysteresis for idealized ``real ocean parameters''~\citep{wang_effect_2010, yuan_hysteresis_2011, song_hysteresis_2019, mei_influence_2019, yuan_dynamics_2019}. A question of whether this idealized geometry applies to realistic western boundary currents remained. In particular, whether the orientation of the gap and the upstream jet pointing into the gap may completely 
change the current system behavior. In this paper we demonstrate, that in application to the LC in the Gulf of Mexico,
the original conclusions do hold and additionally describe peculiarities of the more realistic steady states 
that have implications for LC modeling and forecasting.      

\section{Model}
 
\begin{figure}[hbt]
\centering
\includegraphics[trim=70 30 70 70,clip,width=0.48\textwidth]{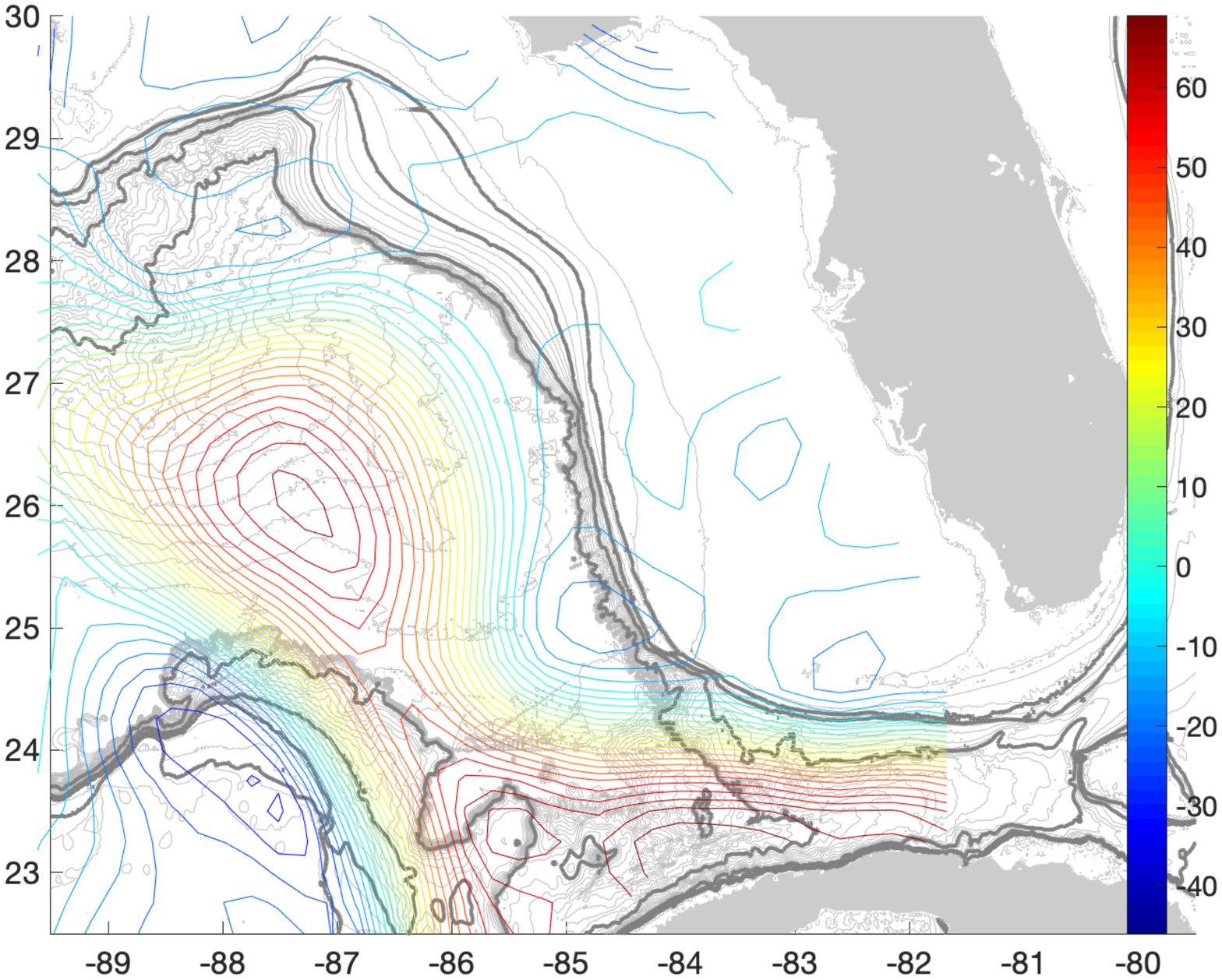} 
\includegraphics[trim=70 30 70 70,clip,width=0.48\textwidth]{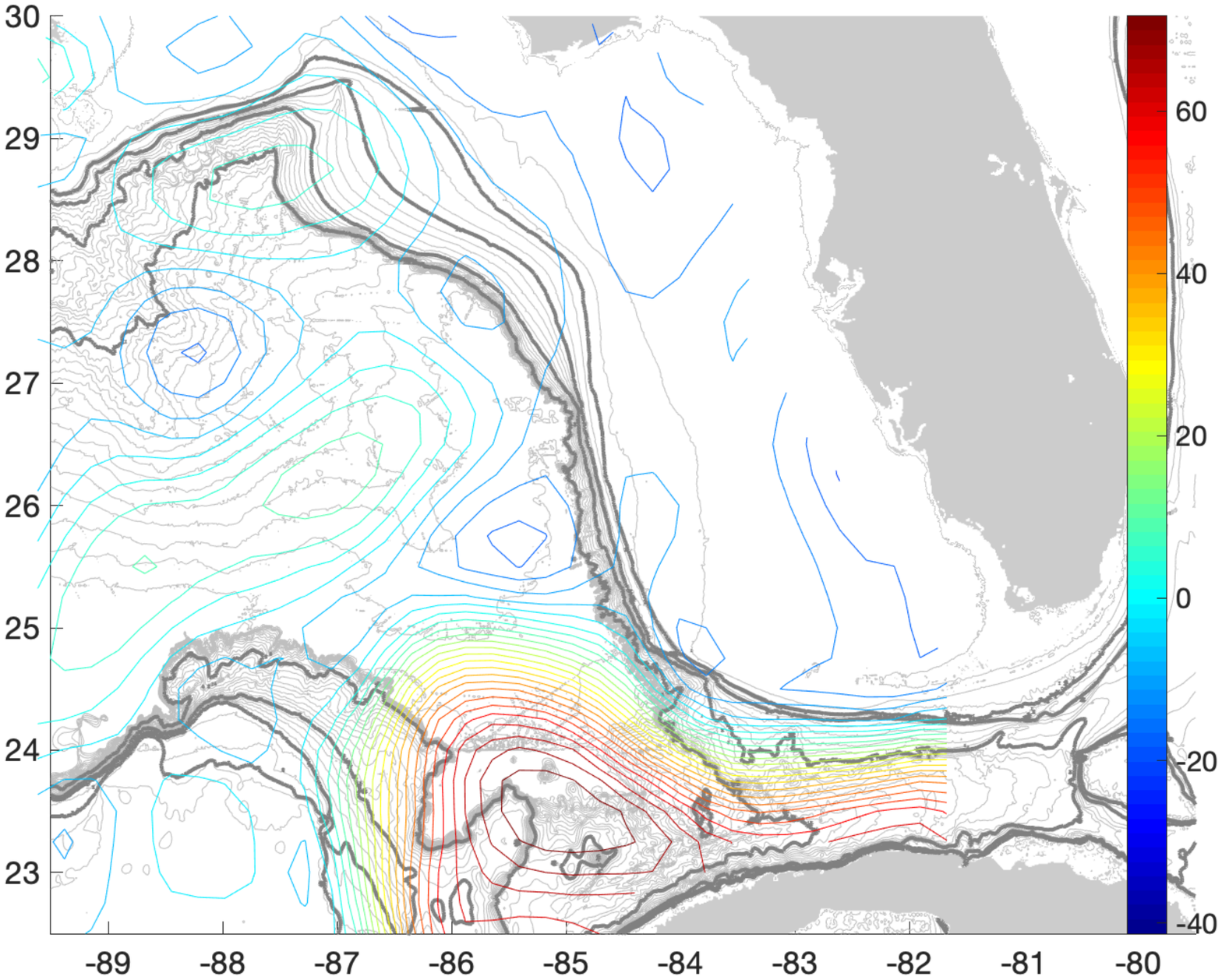} 
\includegraphics[trim=70 30 70 70,clip,width=0.48\textwidth]{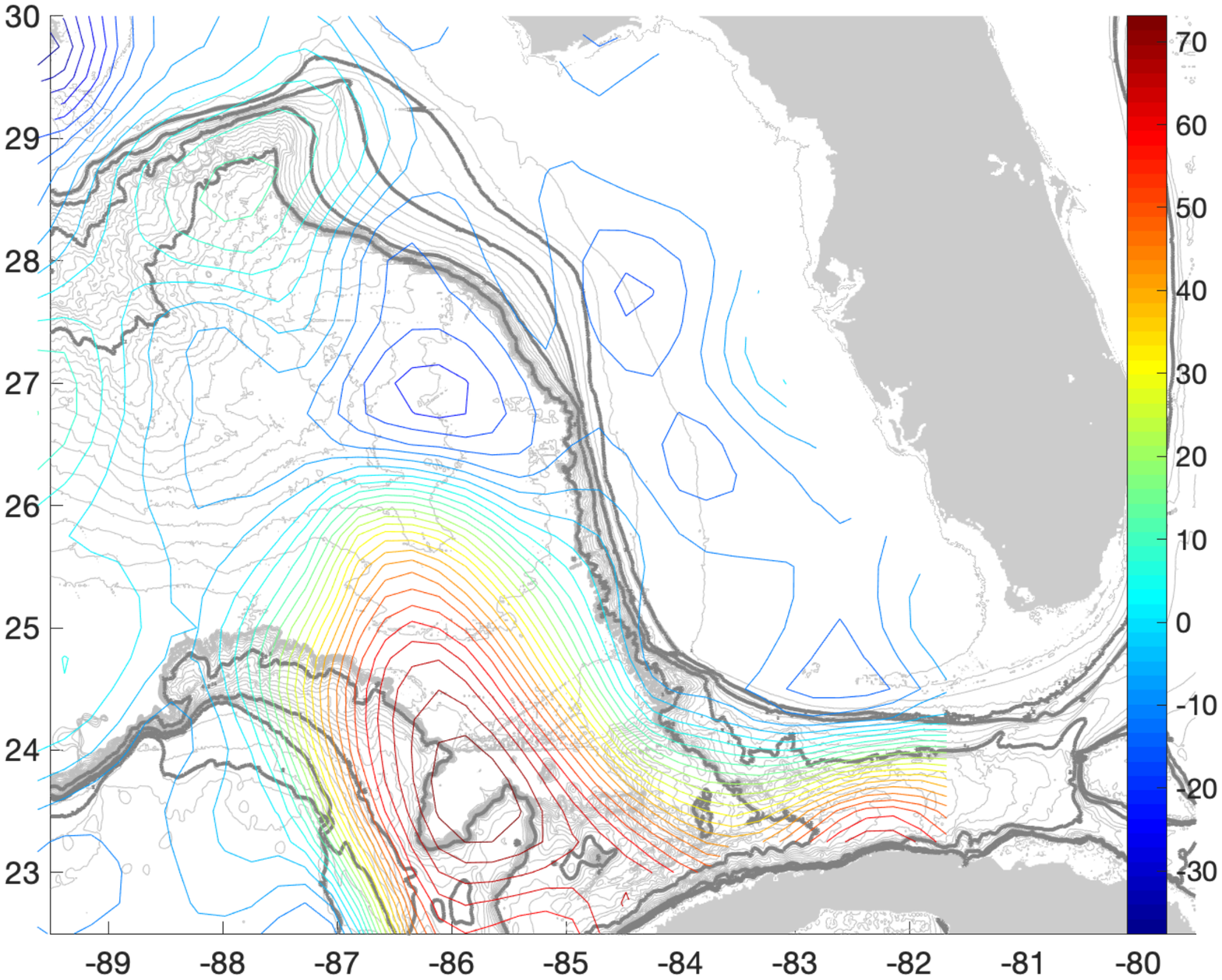} 
\caption{Gulf of Mexico topography with light grey indicating isobaths 5:100:3500m and dark grey indicating isobaths 250,  500, 1000, 2000m.   Grey mask indicates land.   Superimposed are sea surface height contours for: Upper Left) on and around the 193rd day of 2015.  Upper Right) on and around the 57th day of 2017.  Lower) on and around the 154th day of 2017.  Colorbar units are in $cm$. } \label{F1}
\end{figure}

In order to understand the role of the lateral boundary constraints on LC pathways
the potential vorticity advection-diffusion and Poisson equations 
\begin{equation}
J(\psi,q)=\nu \nabla^2 \omega, \qquad  \nabla^2 \psi = \omega
\label{E6:NLP}
\end{equation}
\noindent
are solved on a $\beta$-plane for realistic Gulf of Mexico geometry, where $q=\beta y+\omega$ is the potential vorticity, 
$\beta$ is the beta effect, $y$ is the northward distance, $\omega$ is the relative vorticity, $\psi$ is the flow stream function. The lateral boundary is taken to be the 250m isobath where the bottom rapidly drops from the shelf to the deep continental slope (Figure \ref{F1}). We assume only the planetary beta effect by considering the swift flow, roughly in the top 500m layer, of a fixed thickness, the portion of the flow which is largely shielded from the deep bathymetry by the main thermocline. The wind forcing enters the problem via the inflow boundary condition as an integral effect over the subtropical gyre.

The physical problem has two governing parameters: 
the western boundary current inertial length scale $L_I = \sqrt{U/\beta}$ 
and the Munk viscous layer width $L_M = (\nu/\beta)^{1/3}$ 
dependent on the coefficient of lateral turbulent diffusion $\nu$,
where $U=Q/(H L)$ is the zonal velocity scale in the Sverdrup interior, 
expressed in terms of the total subtropical gyre transport $Q$, 
thermocline depth $H$ and the subtropical gyre meridional scale $L$.
The corresponding nondimensional parameters are $\lambda_I = L_I / L$ and $\lambda_M = L_M / L$.
However, the structure of the western boundary current depends 
on the boundary layer Reynolds number $R=U L_I/\nu=(L_I/L_M)^3$. 

The numerical model utilized in this study was developed to support a series of rotating table laboratory experiments.  Both the experimental setup and validation of the numerical model have been well-established in the literature \citep{sheremet_gap-leaping_2007, kuehl_identification_2009, kuehl_two-layer_2014, mcmahon_viscous_2020}, so only a brief summary is provided here. In nondimensional form, (\ref{E6:NLP}) has $\lambda_M$ instead of $\nu$ and the potential vorticity becomes
$q=y+\delta_I^2\omega$. The domain, as partially seen in figures \ref{F1} and \ref{F2}, spans the Gulf of Mexico and part of the western Atlantic Ocean (77 - 98 west longitude and 13 - 30 north latitude). The kinematic conditions for solving the elliptic equation are $\psi=0$ along all points on the North American continent and the southern boundary. 
The inflow is prescribed as $\psi=\Psi_B(y)$ along the eastern boundary, 
varying from 0 at the southeastern corner to 1 at Cuba (corresponding to the dimensional value of the upstream western boundary current transport $Q$), and remaining 1 to the northeastern corner. The outflow through the northern boundary east of Florida is specified with the Neumann condition. The dynamical conditions are no-slip, zero velocity, at all land boundaries. The values of vorticity at the solid boundaries were calculated assuming the antisymmetry of the tangential velocity component as it was extended outside the fluid domain, which reduces to the formula by \cite{thom_1933} for a straight wall.

The numerical problem is solved using standard finite differences on a rectangular grid 
dividing the domain into $N_x \times N_y$ cells. 
For small boundary layer Reynolds numbers $R=(\lambda_I/\lambda_M)^3$ simple explicit iterations with treating the nonlinear terms as perturbations work well, but for the moderate $R$ the iterations fail to converge. In this case Newton's method has be to employed for finding steady solutions. We consider a state vector $X=(\omega,\psi)$ consisting of values at all grid nodes including the boundaries, the size of this vector is $M=(N_x+1)*(N_y+1)*2$. Substituting an initial guess $X_0$ into (\ref{E6:NLP}) results in the vector of residuals $F(X_0)$ at each grid node of the same size $M$. In order to find the next iteration $X_1$ that brings residual closer to vanishing $F(X)=0$, we need to calculate the Jacobian matrix $J_F[X_0]$ (of size $M \times M$ which depends on $X_0$) of all first-order partial derivatives of $F$ with respect to $X$ and then solve the linear system
\begin{equation}
J_F[X_0] (X_1 - X_0) = - F(X_0)
\label{E6:NWT}
\end{equation}

\noindent
The iterations $X_i$ then continue until the residual vanishes. The iterations were stopped when the residual function (at each node and the overall standard deviation) reduced to and plateaued at $O(-12)$.  It usually sufficed to take 5-7 steps for that. The elements of the Jacobian matrix can be calculated analytically by considering the variational problem (nondimensional) corresponding to~\ref{E6:NLP}.
\begin{equation}
J(\delta\psi,q)+ J(\psi,\delta q) - \lambda_M^3 \nabla ^2 \delta \omega = 0, \qquad               
-\nabla^2 \delta \psi - \delta \omega = 0,                              
\label{E6:VARNLP}
\end{equation}
The variations of the boundary conditions are trivial. Finite difference approximations result in a sparse banded type of $J_F$, and the grids of size upto $1000 \times 1000$ can be solved on a computer with 24~GiB of operational memory. For the present calculation we used the horizontal spacing of $1/20$ of a degree of arc in both directions or $5.55 km$.   

\section{Results}
Steady states of the nonlinear system are found with Newton's method applied to the discrete finite-difference approximation of the equations. This is particularly important, as one can then obtain flow patterns that are not obscured by the external variability (i.e. we are able to identify the underlying dynamics governing the LC). Furthermore, we can determine if distinctly different equilibrium flow patterns are possible for the same values of governing parameters such as the western boundary current transport, as was originally predicted by \cite{sheremet_hysteresis_2001}. Indeed, our investigation shows that for the LC restricted by the geometry of Yucatan and Florida, there exist 5 distinct equilibria (3 stable and 2 unstable) relevant for LC predictability and forecasting. The behavior of the system is summarized in Figure \ref{F2}. We fixed the turbulent viscosity to a typical value $\nu=450$ $m^2 s^{-1}$ used in other numerical models to produce realistic western boundary currents: $300$ $m^2 s^{-1}$ in tropics and $600 m^2 s^{-1}$ in northern parts of the Gulf Stream \citep{large_2001}. The Reynolds number, $R$, then was varied, thus varying the boundary current transport since  $R \sim Q^{3/2}$.  For realistic western boundary currents the inertial effects must dominate over friction, $R>1$.  In our calculations $R=10$ corresponds to a typical western boundary current inertial width of $L_I \approx 60km$. 
That value comes from the Gulf Stream transport of $30$ Sverdrups or $30 \times 10^6 m^3 s^{-1}$ 
spread meridionally over $L=1112 km$ and vertically over the $e$-folding depth scale of $360 m$. 
With $\beta=2.09\cdot10^{-11} (m\cdot s)^{-1}$ for latitude of $24$N, the Sverdrup interior velocity scale is $U=7.5 cm s^{-1}$.  

The LC can be conveniently characterized by its extent $X_p$, which we define as the farthest penetration distance of the streamline $\psi=1/2$ into the Gulf of Mexico from Cuba (shown by red line in panels A and C). The maximum value of stream function $\psi_{max}$ is also helpful for characterizing different states. The stream function is scaled by the upstream western boundary current transport $Q$. All the flow that enters the computational domain in the Carribean Sea (the unit transport) must exit as the Gulf Stream northward along Florida.  However, the $\psi_{max}$ can exceed unity as the LC drives the recirculation eddy inside the Gulf of Mexico (Panels B-D).
The tracing of $X_p$ and $\psi_{max}$ for the steady solutions are shown in the last panels at lower right as a function of $R$. The letters A through G along the $X_p$ curve (red) correspond to the flow patterns shown in the corresponding Panels. 

If we start from the linear solution, $R=0$, no-advection, (Panel A) the streamlines are blocked by Yucatan and do not penetrate westward. As the advection increases, the LC rapidly penetrates into the Gulf of Mexico reaching the northern coast at $R=2.5$.
After that the flow pattern (like the one shown in Panel B for $R=10$) changes little until $R \simeq 30$. Then the intensity of the LC eddy increases substantially. Near $R\simeq 40$ the LC eddy occupies nearly the whole area between Yucatan and Florida and its transport increases by the factor 1.4-2. There are actually six steady solutions possible for the same $R$ in that region. They have similar patterns (as shown in Panel C) but differing by $\psi_{max}$ and by the points at which the gyre touches the coast.

Further tracing the solution branches, takes us back to smaller $R$ and a smaller LC eddy (Panel D). There are three folds near $R=10$ and correspondingly four more steady solutions marked by D,E,F,G each representing a different extent of the LC. Finally, in the state G the LC assumes its retracted state and remains such as $R$ increases beyond. 

\section{Discussion}

\begin{figure}[hbt]
\centering
\includegraphics[trim=0 0 0 0,clip,width=01.0\textwidth]{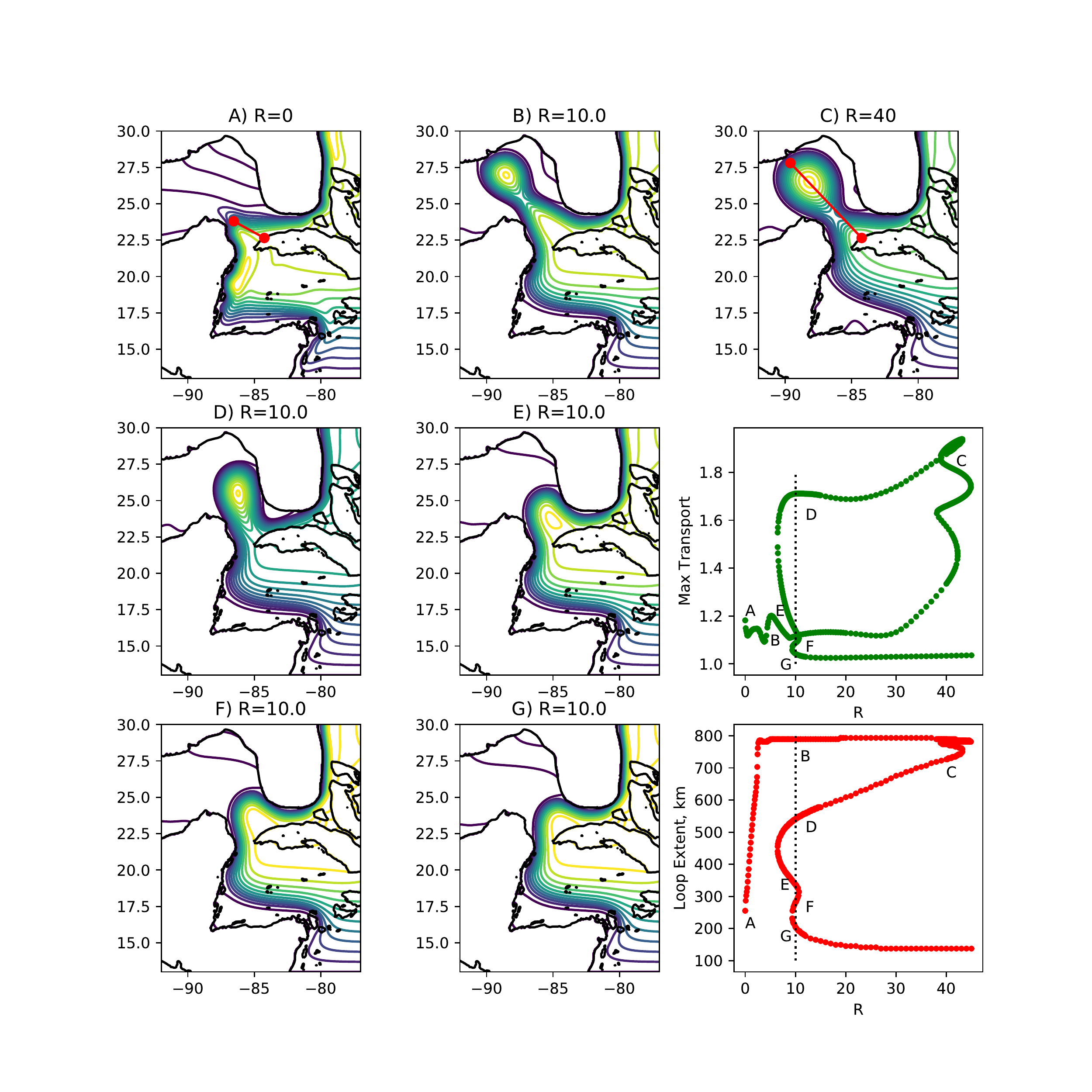} 
\caption{The streamfunction patterns, the maximum streamfunction, and the LC extent illustrating tracing of steady solutions
as the boundary Reynold number, $R$, is varied. The panels B,E,G (stable sates) and panels D,F (unstable states)
represent five solutions for exactly same parameter $R=10$. The panel/state letters are marked along the LC Extent vs $R$ (red) plot in the lower right. The definition of LC extent is marked by red line and red circles in panels A and C. The streamfunction contour interval is 0.1 of the total transport. The land mask was generated based on the 250m isobath, except Cayman, Jamaica and small islands in the Carribean were ignored in order to specify the zonal inflow. }\label{F2}
\end{figure}

%

Around $R=10$ (which correspond to typical oceanographic conditions) there are 5 steady solutions (B,D,E,F,G) possible in the system.  According to the alternating stability of branches in bifurcation theory, the equilibria B,E,G (increasing $R$) are stable, while D and F (decreasing $R$) are unstable; see \cite{ierley_multiple_1995, Sheremet_OM_2002} for further details. 
We emphasize that we mean here the stability with respect to only one, but very significant, eigenvalue whose imaginary part is zero (a non-oscillatory eigen mode). At the folds of the solution branches, the real part of this eigenvalue (representing the exponential decay or amplification of the mode) also vanishes, the eigenvalue becomes zero, which gives rise to the non-uniqueness of the steady solutions  and multiple steady states.  We note that there may be other modes affecting the linear stability, but those are oscillatory and may lead to Hopf bifurcations. Conducting a full stability analysis is beyond the scope of this  
paper, which will be done in a subsequent work. 

The stability type of the states is extremely important to Loop Current predictability/forecasting, as the unstable states D and F act as a separatrix between the stable basins of attraction corresponding to solutions B, E, G.  This can be understood by considering the simpler case of vibrating system analysis.  In vibrating systems analysis, the state and evolution of the system are often cast into a state space (or phase space) diagram.  This diagram is constructed by first identifying steady state solutions of the system and then identifying their stability type.  The system is repelled from
  unstable solutions and attracted to stable solutions.  Thus, the system evolution in the state space can be identified with unstable solutions delineating the basins of attraction of the stable solutions.  Here, we view the LC as a dynamical system, similar to \cite{lugo-fernandez_is_2007}, and we can interpret the unstable states as the higher-dimensional separatrices of our system.  That is, flow in the state D, if perturbed, will evolve toward B or E depending on the sign and structure of perturbation.  Similarly, the flow in state F will tend to evolve toward E or G.  Thus, knowing the structure of unstable LC states, allows for insight into the transitional dynamics between the stable LC states.  

\begin{figure}[hbt]
\centering
\includegraphics[trim=0 0 0 0,clip,width=01.0\textwidth]{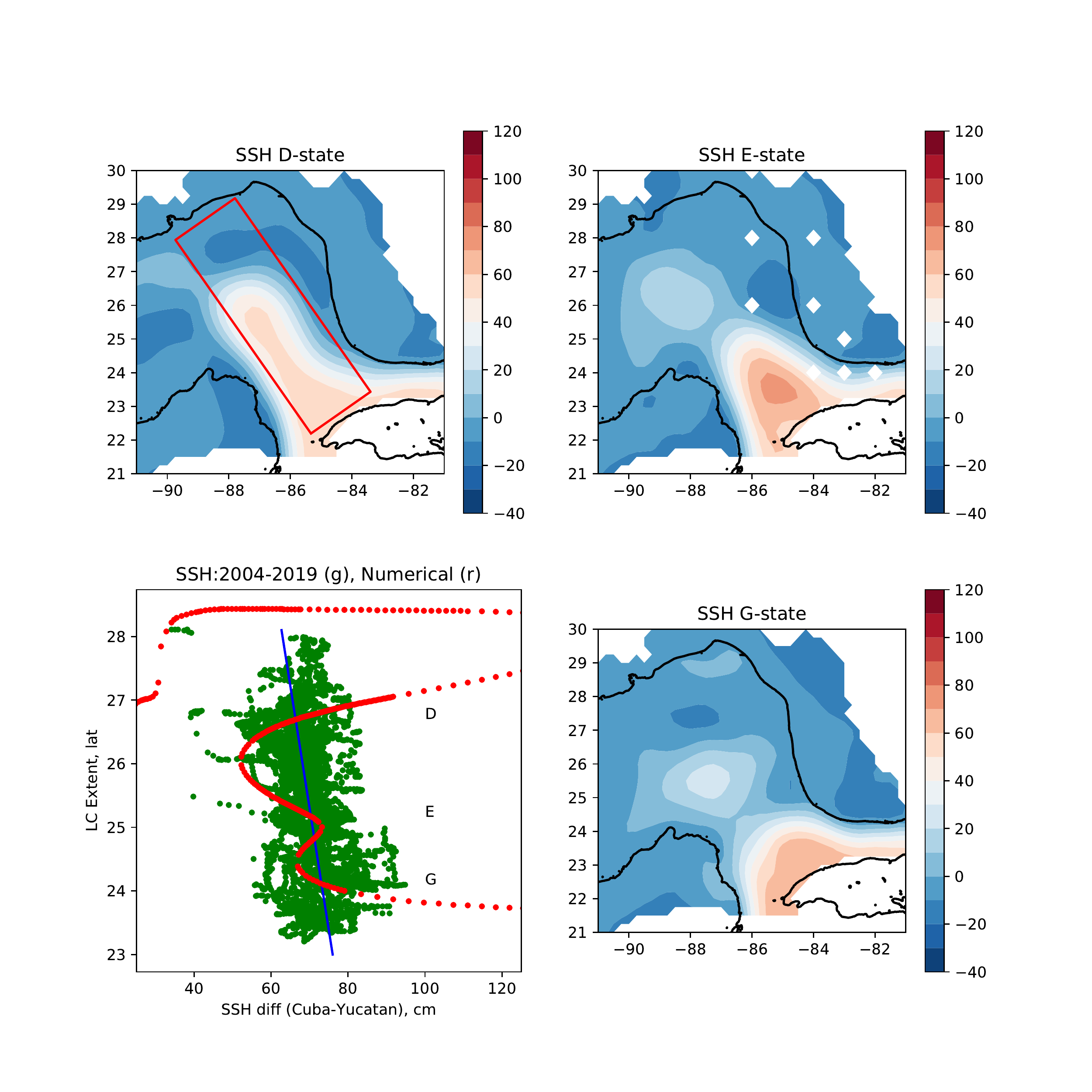} 
\caption{The LC Extent (lower left panel): from the satellite altimetry SSH fields (green dots) and a linear fit (blue line); the red dots are LC Extent from steady numerical solutions. 
The statistically averaged SSH fields corresponding to the LC near the states D,E,G are shown the surrounding panels, the SSH units are cm. The red rectangle in the upper left panel shows the swath where the 35cm SSH contour was searched.} 
\label{F3}
\end{figure}

To validate our approach, we compared our results to the satellite altimeter Sea Surface Height (SSH) dataset for the Gulf of Mexico. The dataset is available from GCOOS (URL https://geo.gcoos.org/ssh/data1/) and now is spanning the years 2004-2019 (we excluded the year 2020 since it currently only spans January and February). Details of the data and its processing can be found on the GCOOS website and in \cite{Leben_SSH}.  We measured the extent of the LC as a distance from the point (-84.250E, 22.666N) at the northwestern coast of Cuba to     
the point along the 240km wide swath directed along the azimuth of -35 degrees from true north where the isoline of half LC transport is located.  In the SSH field that corresponds to the isoline of 35cm, half the difference between the values at Cuba and Yucatan or between Cuba and Florida, which on average is about 70cm. This SSH difference is a proxy for the total geostrophic transport of the current. The LC extent is expressed in terms of the latitude of that point (Figure \ref{F3} lower left Panel, green dots; the blue line is a linear fit). The same algorithm for LC extent is applied to our steady numerical solutions and the data are superimposed as red dots; we matched the mean value of $\Delta$ SSH=70cm to correspond to the realistic numerical case with R=10 discussed above. 

Most of SSH derived data points fall in the range of LC latitudes between our states D and G and correspond to similar transports. There is a small but noticeable decrease of LC Extent with increased transport of the current 
(note the linear fit, blue line). Next we performed statistical averaging of SSH fields by selecting only the fields with the LC extent within a small range (plus or minus 0.1 degrees latitude) around the intersection points of blue and red curves, 
and thus obtained averages corresponding to states D, E and G illustrated in the other panels of Figure \ref{F3}.
They indeed have remarkable resemblance to the numerically calculated patterns of Figure \ref{F2} (note the same arrangement
for ease of comparison). The state G is a stable fully retracted LC, E is also a stable state with a medium extent.
The state D is unstable, and, as we mentioned earlier, it acts as a boundary of the stable basin. If LC surpasses that extent,
it will tend to evolve towards state B, continue to grow and in real ocean will pinch off a westward drifting LC eddy. 
That is why the state D is not realized in the SSH observations.

It is quite interesting that our analysis identifies 3 stable LC states, when the traditional view has always been that of a 2 state system: extended and retracted.  Though this should not be too surprising given the difficulty in identifying `true Loop Current states' from time-dependent observational or numerical datasets \citep{kuehl_application_2014, weisberg_loop_2017}.  However, the stable flow patterns identified do have precedent in the observational data records.  Figure \ref{F1} identifies sea surface height fields from the historical satellite altimetry data that roughly correspond to the stable LC states identified in Figure \ref{F2}:  Figure \ref{F1} upper left panel is a looping state that was present on and around the 193rd day of 2015 and corresponds to state B.  Figure \ref{F1} upper right panel  is a leaping state that was present on and around the 57th day of 2017 and corresponds to state G.  Figure \ref{F1} lower panel  is an
  intermediate state that was present on and around the 154th day of 2017 and corresponds to state E.  Note, while we can not guarantee that these observational flow patterns are `true Loop Current states', all flow fields identified persisted for approximately 2 months.  

\begin{figure}[hbt]
\centering
\includegraphics[trim=0 0 0 0,clip,width=01.0\textwidth]{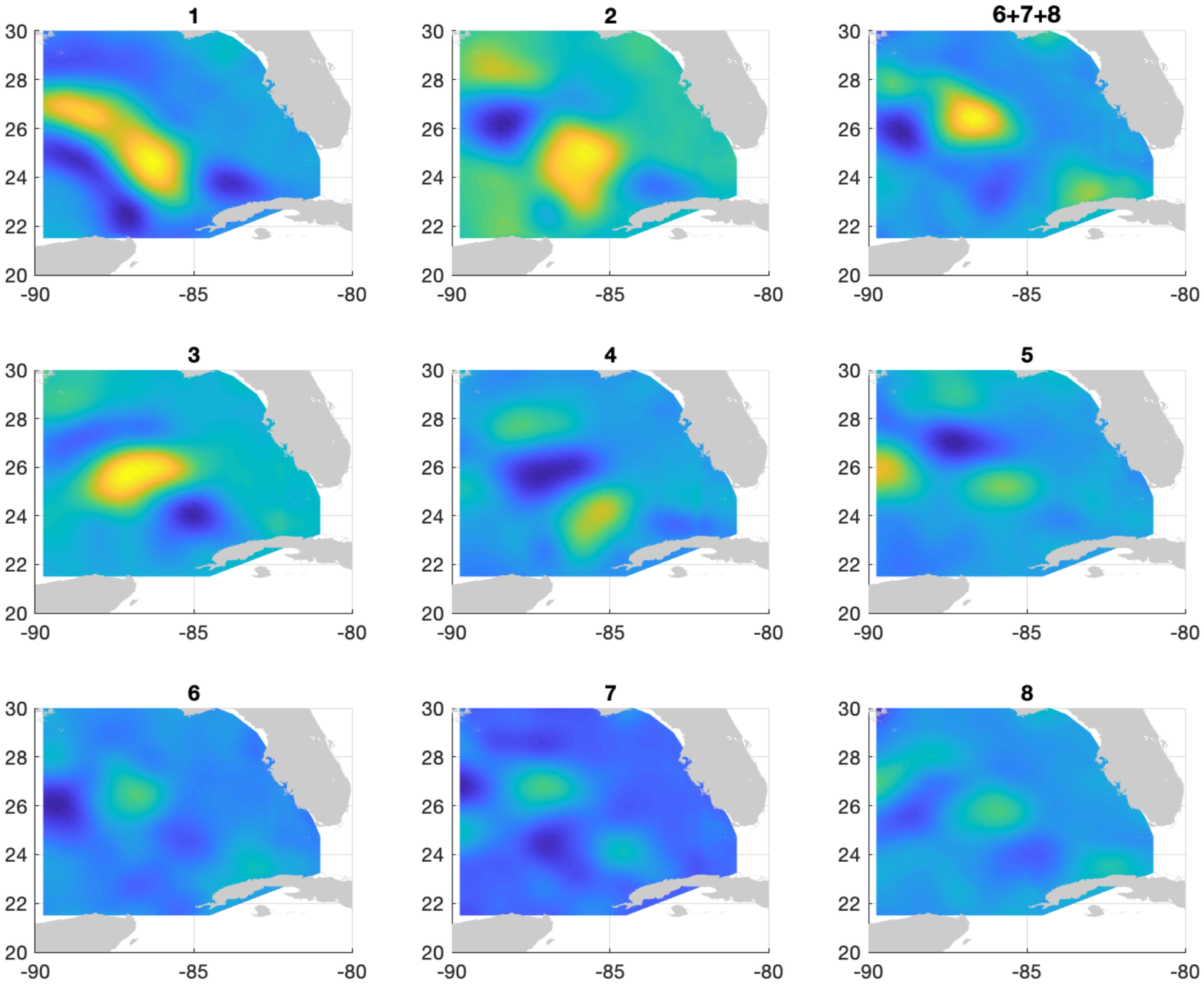} 
\caption{SOD modes shapes calculated from SSH observational data.  Mode are identified by mode number,  the most dominant being 1 and reducing as mode number increases.  Grey identifies dry land. } 
\label{F:SOD}
\end{figure}

To further investigate the SSH observational data, a Smoothed Orthogonal Decomposition (SOD) \citep{chelidze_smooth_2006, khan_toward_2020} investigation was conducted.  SOD considers the constrained maximization problem 

\begin{equation}
\underset{\phi}{max} \| X \phi\|  \hspace{1cm} \textrm{subject to} \hspace{1cm} \underset{\phi}{min} \|V \phi\|,
\end{equation}

\noindent
where $V$ is the temporal derivative of $X$.  The corrsponding Rayleigh's quotient becomes

\begin{equation}
\underset{\phi}{max} \left\{ \lambda(\phi) = \frac{\|X \phi\|}{\|V \phi \|} \right\} \rightarrow \underset{\phi}{max} \left\{ \lambda(\phi) = \frac{\phi^T \Sigma_x \phi}{\phi^T \Sigma_v \phi} \right\},
\end{equation}

\noindent
where $\Sigma_v$ is the covariance matrix of V.  Variational techniques may be applied and result in a generalized singular value decomposition (GSVD) problem for $X$ and $V$ or correspondingly the generalized eigenvalue problem $\Sigma_x \phi_i = \lambda_i \Sigma_v \phi_i$.  It follows that SOD considers those modes which maximize amplitude variance projection, while at the same time being as smooth in time a possible.  As nature tends to behave in a ``smooth'' fashion, this ``smoothness'' property has been shown to be more physically relevant than the amplitude based Proper Orthogonal Decomposition (POD, or EOF, or PCA) analysis \citep{Chelidze_Slow}.  In the case of the Gulf of Mexico, we might expect the dominant LC states to be separated by time-scale which is indeed what is observed (Figure \ref{F:SOD}).  The top row of the figure identifies (from left to right) the dominant signal of an extended LC state, an intermediate LC state, and a retracted LC state.  The middle row panels identify a periodic eddy-shedding LC state, and the bottom row panels illustrate oscillatory behavior of the retracted state.  These results, in combination with numerical work presented above, strongly suggest the presence of at least three dominant LC states in the Gulf of Mexico, in addition to a periodic eddy-shedding states similar to that identified by \cite{kuehl_two-layer_2014}.

The identified states are also consistent with our current understanding of the governing vorticity dynamics of LC systems and their structure highlights the importance of current-boundary interaction.  The LC system essentially 
involves a combination of: 1) a boundary current separation problem.  2) a current reattachment problem. 3) a vorticity budget closure problem.  

The question of flow separation from a curved boundary (Yucatan Peninsula) is among the most difficult in fluid mechanics. 
A criterion based on the cancellation of the inviscid flow singularity at the point of separation is known as Brillouin-Villat condition after \cite{Brillouin_1911, Villat_1914} who proposed it. 
The analogy of the gap leaping problem and the teapot effect was discussed in ~\cite{sheremet_hysteresis_2001}.
The separation of pouring flow from the teapot spout was investigated by \cite{VandenBroeckKeller_1989} 
who constructed ideal fluid solutions with separation from arbitrary points downstream of the lip
and concluded that it is the conditions in the far field downstream that are crucial. 
Thus, in the gap leaping problem it follows that the position of flow reattachment (to the west Florida coast)
exerts a more significance control over the LC state than the separation point (i.e.~the reattachment point controls the separation point, not vice-versa). The present calculations, with viscosity, suggest that the range of movement of the reattachment point is much greater in comparison.


We did investigate the sensitivity of results to boundary conditions by varying the coefficient of proportionality $k$ in the formula connecting the boundary vorticity $\zeta_0$ to the nearby tangential velocity $(\psi_1-\psi_0)/h$:  $\zeta_0=k*(\psi_1-\psi_0)/h^2$, where the values $\zeta_0$ and $\psi_0$ are at the boundary and $\psi_1$ is the value of the stream function one grid step $h$ away from the  boundary.  The standard case (no-slip) reported here (\cite{thom_1933}, $k=2$) and the partial slip ($k=1$) give very close flow patterns since they produce realistic flow separation from Yucatan and reattachment at Florida. In fact, the pattern corresponding to state E can be maintained for $k$ as small as 0.43. In contrast, if the slip condition (no stress, $k=0$) were specified, it would prevent the current from separating and it would follow the coastline around Yucatan into the western Gulf of Mexico.

The vorticity dynamics of LC systems have been thoroughly addressed by the series of works by Sheremet and Kuehl (referenced above), and recalling \cite{pichevin_momentum_1997}, can be applied to understand dynamical balances of each state.  
\begin{itemize}
\item The low momentum branch B, has insufficient inertia to separate from the Yucatan topography before 24.5 north latitude.  However, unlike state A, it does have enough momentum to overshot the latitude of southern Florida and must eventually make a right turn as describe originally by \cite{Reid_1972}.   This state would normally result in a periodic eddy shedding state if not for the northern Gulf of Mexico Slope.  The vorticity advection into the LC system can be balanced by a periodic eddy shedding state \citep{cannon, kuehl_two-layer_2014} or in the steady state through interactions with northern Gulf of Mexico topography.  
\item The high momentum branch G, has sufficient inertia to separate early from the Yucatan topography.  The effective radius of curvature of the current flowing directly from the Yucatan to the Florida Straits is larger than the distance between the two coastlines and thus the current traverses this gap fairly undisturbed.
\item The middle branch E is a balance between the the low and high inertial branches.  The current has sufficient inertial to separate from the Yucatan topography and leap to the west Florida coast.  However, increased contact/interactions with the west Florida coast are required to dissipate excess vorticity advected into the Gulf of Mexico, very similar to that documented in \cite{kuehl_two-layer_2014}.
\end{itemize}

Note, it seems counterintuitive that the flow, as the intertia is increaded, will leap from Yucatan to Florida, 
despite the fact the the jet initially points to the interior of the Gulf. The inertial flow due to the $\beta$-effect has a characteristic width $L_I$. If this scale becomes comparable with or larger than the gap half width, then the current cannot squeeze into the Gulf of Mexico, in other words, the incoming and outgoing boundary current would merely not fit through the gap (see the curvature based transition in \cite{kuehl_two-layer_2014}).      

\noindent
Conclusions:
We identified multiple equilibrium states of the Gulf of Mexico Loop Current 
by solving with Newton's method the nonlinear steady potential vorticity advection-diffusion equation restricted by the realistic lateral boundaries. This approach allows us to find both the stable and unstable states (flow patterns) of the system
that are dynamically balanced. In contrast, other approaches such as modal decomposition (Empirical Orthogonal Functions, Singular Value Decomposition, Dynamic Mode Decomposition, etc.), machine learning, self-organizing maps, etc. are based on observational (or numerical) data correlations.  We intentionally restricted the physics to the quasigeostophic dynamics to capture the most important mechanism of the LC formation: the balance between the planetary $\beta$-effect that promotes the LC penetration into the Gulf of Mexico and the inertia that promotes the current to leap directly from Yucatan to the Florida Straits. We are currently working on generalizing this approach to include additional physical effects: topographic $\beta$-effect, stratification, turbulent viscosity model
in the framework of realistic numerical models with a goal of using the obtained states as a basis for analysis of the system predictability.

\acknowledgments
The authors are thankful to the National Science Foundation, USA for funding this research via grant number 1823452, and to the National Academies of Sciences, Engineering and Medicine (NASEM) UGOS-1 via grant number 2000009918 and UGOS-2 via grant number 200011071.  The authors are also grateful to two anonymous reviewers for their valuable comments and suggestions for improving the manuscript. The numerical code used in this work can be found at http://sites.udel.edu/kuehl-group/software/ .






%
%
%
\bibliographystyle{ametsoc2014}
\bibliography{Bib_R5}

%

%

\end{document}